RESEARCH ARTICLE

# Connectivity jamming game for physical layer attack in peer to peer networks

Ying Liu* , Andrey Garnarv and Wade Trappe

Wireless Information Network Laboratory, Rutgers University, North Brunswick, USA

## ABSTRACT

Because of the open access nature of wireless communications, wireless networks can suffer from malicious activity, such as jamming attacks, aimed at undermining the network's ability to sustain communication links and acceptable throughput. One important consideration when designing networks is to appropriately tune the network topology and its connectivity so as to support the communication needs of those participating in the network. This paper examines the problem of interference attacks that are intended to harm connectivity and throughput, and illustrates the method of mapping network performance parameters into the metric of topographic connectivity. Specifically, this paper arrives at anti-jamming strategies aimed at coping with interference attacks through a unified stochastic game. In such a framework, an entity trying to protect a network faces a dilemma: (i) the underlying motivations for the adversary can be quite varied, which depends largely on the network's characteristics such as power and distance; (ii) the metrics for such an attack can be incomparable (e.g., network connectivity and total throughput). To deal with the problem of such incomparable metrics, this paper proposes using the attack's expected duration as a unifying metric to compare distinct attack metrics because a longer-duration of unsuccessful attack assumes a higher cost. Based on this common metric, a mechanism of maxmin selection for an attack prevention strategy is suggested. Copyright © 2017 John Wiley & Sons, Ltd.

## 1. INTRODUCTION

The connection properties of a wireless network can degrade easily with adverse environments, such as a tall building that obstruct signals or strong noise that interferes with normal communications. Links breaking is a common phenomena in wireless communications. However, if a malicious jammer purposely breaks a link and separates a node from a network, this harmful behavior can seriously interrupt the normal operation of the network, especially if the node happens to be the hub of several routes. Therefore, investigating the impact of the removal of critical nodes and analyzing the jammer's strategy in choosing a node for an attack is essential to maintain network connectivity in adversarial settings.

Many techniques have been presented to detect the intrusive behavior of an attacker. In [1], a survey of intrusion detection techniques is given. In [2], several host-based and network-based intrusion detection systems are surveyed as well as their characteristics are described. In [3], a Bayesian approach was used to detect an intruder in a spectrum band while taking into account whether the intruder sneaks for file-downloading or streaming video. In [4], a Bayesian learning mechanism is used to design a scanning strategy if there is incomplete knowledge whether the intruder is present. In [5], fictitious play from game theory is adopted to classify the type of a jammer based on the historical belief in the throughput under attack uncertainty. In [6] and [7], data mining techniques to recognize anomalies as well as known intrusions are presented. In [8], a lightweight and generic localization algorithm is developed for finding the location of a jamming device after detecting its malicious activity. In [9], an algorithm of localization in peer to peer networks based on *Q*-learning approach is suggested. However, none of these papers considered the intruder's impact on network connectivity nor mechanisms that can maintain connectivity in the presence of such an adversary.

In this paper, different anti-jamming strategies versus jamming attacks aimed at harming different network

     

characteristics, like connectivity or throughput, are investigated in a uniform framework. In such a situation, an entity intending to protect a network faces a problem that while an adversary might apply a fixed set of jamming tools, the underlying intent or strategy behind an attack can be quite varied, depending largely upon the network's characteristics. In particular, the metrics associated with such an attack can be incomparable (e.g., network's connectivity and total throughput). To deal with the incomparable metrics problem, this paper makes the following contributions:

(1) A general stochastic game involving the protection of a network, where a jammer might sense nodes' scanning and switch to a hiding (i.e., silent) mode, is suggested. In the considered model, the meaning of the instantaneous costs depends on the type of jamming attack strategy being applied. For the attack aimed at harming network connectivity, the instantaneous costs for the jammer are described by the algebraic connectivity or Fiedler value of the network [10,11]. For the attack aimed at harming throughput, the costs are the network's throughput. It is important to note that because the network protector aims to maximize cost of an adversary's attack, this game is fundamentally about the prevention of an attack rather than about the network's protection.

(2) Because the longer an attack is unsuccessful leads to higher cost, we propose a unified metric that makes it possible to compare such attacks whose metrics of success (e.g., harming connectivity or throughput) would otherwise be incompatible. In particular, we propose the use of the attack's expected duration and, based on this common metric, we arrive at a mechanism of maxmin selection for nodes' scanning strategy is suggested.

The paper is organized as follows. In Section 2, related works are discussed. In Section 3, jamming attacks aimed at undermining network connectivity and network throughput are presented, as well as a preliminary overview of algebraic connectivity. In Section 4, the problem of preventing an attack against a network is formulated as a stochastic game, and it is solved explicitly. In Section 5, a tool for comparing defenses against jamming attacks aimed to harm different network characteristics is developed. In Section 6, results of numerical evaluation for the optimal solutions and their dependence on network characteristics are supplied. Finally, in Section 7, conclusions are given.

## 2. RELATED WORKS

Studies that explore the connectivity of networks and their topological properties can be found in the literature and a sample includes[12–14]. The mostly widely adopted approach for summarizing a network's topological connectivity involves the calculation or prediction of node degree from statistical results. Network connectivity data have been collected from a variety of real networks, such as social networks[15] and citation networks[16], and is complemented by mathematical models, such as scale-free networks, where the node degree follows a probability distribution that decays in a power-law, or a Poisson random networks, whose nodal probability distribution follows a Poisson distribution.

An important research area that applies to all of the network models mentioned earlier involves studying network connectivity under malicious attacks. While these attacks can happen at each network layer, most research about network connectivity traditionally focuses on designing secure routing protocols by which packets can route around a black hole or wormhole in networks[17–19]. Those routing protocols usually aim to find the most efficient and free path in a topological graph after an attack happens. On the other hand, the impact of a broken single link or removal of a node in a path, and the resulting diffusion of attack damage across the broader network context has been studied much less, particularly when the connectivity issues appear at the physical layer.

Ensuring the robustness of the physical layer typically involves examining links in isolation (e.g., robust error coding), and notably separate from the broader network context. The robustness of *networks* at the physical layer should examine the network's performance after one node/link, or even several nodes/links, are degraded or removed at the physical layer. For example, an attacker can randomly delete several nodes or strategically delete nodes according to his purposes though targeted interference, aimed at greedily removing nodes with higher degree first or deleting nodes in high density areas in order to exacerbate the damage. To the best of our knowledge, most prior research into the resilience of networks are statistical, and they fail to consider the interaction between legitimate nodes and the attacker and, moreover, tend to consider that the jammer behavior is random in how it eliminates nodes, without a deeper strategy behind how to maximize its attack effectiveness.

Game theory is a natural tool to investigate and rationalize about a jammer's behavior. Game theory investigates the interactions between players to arrive at equilibrium strategies for both sides[20]. In [21], a survey of works that applied game theory to deal with network security at each layer is given. Physical layer security can be described by game theory both in the form of a Nash game and a Stackelberg game. Game theory papers at the physical layer security often model the rational behaviors of a jammer, or an eavesdropper, or cooperative behavior between them, to solve the problem of allocating transmission power or increasing transmission rate. Typically, the utility function being employed is Shannon capacity, signal to interference and noise ratio (SINR), information entropy or bit error equations. There is a limited set of works dealing with maintaining the connectivity of the network topology. In [22], a problem of minimizing the probability that the spanning tree disrupted by an adversary attack was studied. In [23], to identify key players engaged in attacking a network, the Shapley value was applied. In [24], a problem with two types (good and bad) of users was studied by a repeated game, where good users were willing to trade energy for connectivity depending on neighbors' behaviors, while bad users try to destroy connectivity and lure the good users to waste energy. In this paper, we consider the game where users' throughput, and network connectivity are combined in a unified framework.

## 3. FORMULATION OF THE PROBLEM

As a motivating scenario, we consider a zone that involves $n$ nodes (users) allocated at points $(y_{1i}, y_{2i})$, $i \in [1, n]$ and operating in a P2P full duplex communication fashion, which allows nodes to communicate in both directions. Let $e_{i,j}$ be *a duplex communication link* (a channel) for communication between node $i$ and $j$. A possible connection between any two nodes is defined by the channel's condition, the mutual distance between nodes, receiver threshold, transmission power, and transmission protocols, that is, the collision avoidance protocol in Medium Access Control (MAC) layer. Let $\mathbf{h} = \{h_{i,j}\}_{i,j=1}^{n}$ be an $n \times n$ matrix describing the communication capabilities between nodes (fading channel gains). Generally, this matrix might be non-symmetrical, that is, the component $h_{ij}$, mapping communication from node $i$ to node $j$, might be different from the component $h_{ji}$, mapping communication from node $j$ to node $i$. Some components of matrix $\mathbf{h}$ might be zero, reflecting the fact that either communication between these nodes is not allowed, or this node has no intent to communicate with another. For a particular, $h_{ii} = 0$ for any $i$, as there is no need for a node to communicate with itself. Therefore, based on these conditions, the complete possible communication topology is already determined. Let $P_{i,j}$, $i, j \in \{1, \ldots, n\}$ be transmission protocol between nodes, that is, $P_{i,j}$ represents transmission power from sender $i$ to receiver $j$ on channel $e_{i,j}$. Then, by applying a Shannon-type formulation for channel capacity, the throughput for communication node $i$ to node $j$ is $T_{i,j} = \ln\left(1 + h_{i,j}P_{i,j}/(\sigma^2 + \sum_{k=1, k\neq j}^{n} h_{k,j}P_{k,j})\right)$, where $\sigma^2$ is the background noise.

In the zone, besides of the legitimate nodes, an adversary jammer equipped with limited power is present to harm communication. Its location is given by the coordinates $\mathbf{x} = (x_1, x_2)$. The effect of jamming (jammed throughput) for communication node $i$ to node $j$ is $\ln\left(1 + h_{i,j}P_{i,j}/(\sigma^2 + \sum_{k=1, k\neq j}^{n} h_{k,j}P_{k,j} + g_j J/d_j^2)\right)$ which depends on the distance $d_j = \sqrt{(x_1 - y_{j1})^2 + (x_2 - y_{j2})^2}$ between the jammer and the receiver, the fading channel gain $g_j$, and the jamming power $J$ being applied. Because the jammer has a power limitation, he cannot effectively impact the communication of nodes located far away. However, if the attacker is allocated close by a particular node, then the jammer can effectively jam that node from all incoming traffic received. Because of all incoming messages for the jammed node are blocked, the acknowledge messages(ACK) corresponding to his request to establish communication with other nodes are also blocked [25]. Thus, the jammed node cannot recognize its neighboring nodes, and hence it cannot communicate with them. So, when the jammer is attacking a node, it can disrupt bi-lateral communications (incoming and outcoming) for that node. In the meantime, the Request to Send/Clear to Send problem can also make the receiver detecting the existence of a hidden terminal and ceasing the transmission to the target. Ultimately, the jammer achieves his goal by blocking the whole receiving and sending functions of the target. To describe the effects of blocking such bi-lateral communications, we assume the jammer's ability to block the communications is equivalent to its ability to disrupt the communication links in bilaterally.

Note there are lots of types of jamming attacks. A reader can find comprehensive surveys of such threats in [26]. In this research, we introduce a new type of attacks which targets the network's connectivity, and compare it with the jamming attack which targets the network's throughput.

### 3.1. Cost of breaking connectivity attack

In this section, we describe breaking connectivity attack and its cost. This is a jamming attack targeting to break duplex communication links between nodes. In order to break a link from sender to receiver in its physical layer, the received SINR must be smaller than the threshold $\omega$. Let the threshold be the same for all the nodes. Then, we can express the condition of a broken link from node $i$ to node $j$ by

$$\frac{h_{i,j}P_{i,j}}{\sigma^2 + \sum_{k=1, k\neq i}^{n} h_{k,j}P_{k,j} + g_j J/d_j^2} < \omega. \quad (1)$$

Thus, to break communication from node $i$ to node $j$, the following induced jamming power has to be applied

$$\left\lfloor \frac{h_{i,j}P_{i,j}}{\omega} - \sigma^2 - \sum_{k=1, k\neq i}^{n} h_{k,j}P_{k,j} \right\rfloor_+ \leq \frac{g_j J}{d_j^2}, \quad (2)$$

where $\lfloor \xi \rfloor_+ = \max\{\xi, 0\}$. Because of block ACK to break all the bilateral links on node $j$ from other nodes, the following induced jamming power has to be applied:

$$\max_{i, i\neq j} \left\lfloor \frac{h_{i,j}P_{i,j}}{\omega} - \sigma^2 - \sum_{k=1, k\neq i}^{n} h_{k,j}P_{k,j} \right\rfloor_+ \leq \frac{g_j J}{d_j^2}. \quad (3)$$

This condition can be achieved by having sufficient closely positioned adversary to node $j$ in spite of its limited jamming resource. Such adversary's strategy allows the elimination of a selected node from networks communication to cause network disruption in terms of connectivity.

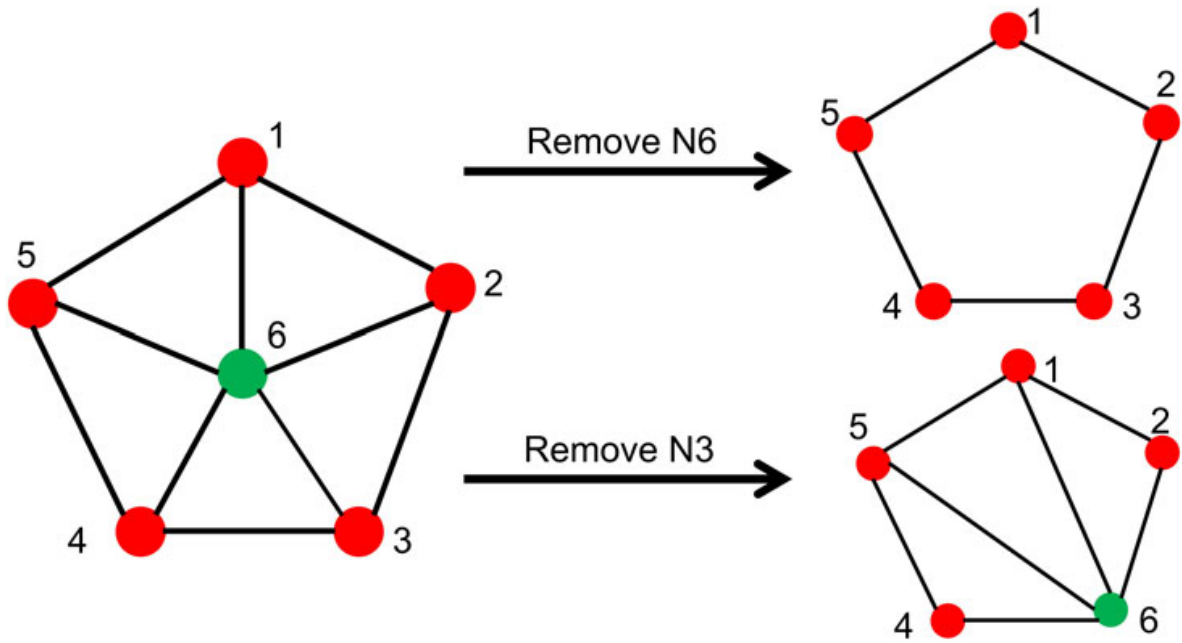

**Figure 1.** Fiedler values of different remaining graph are comparable when the number of nodes is the same.

To deal with the remaining network connectivity, a concept of Fiedler value[27] developed in spectral graph theory can be applied.

Fiedler value is the second smallest eigenvalue of the Laplacian matrix, $L(V, E)$, of a network's topological graph, $\Gamma(V, E)$ where $V$ is a vertex set and $E$ is the edge set connecting two vertices in the graph. The Fiedler value is always non-negative, and its amplitude is proportional to the graph connectivity. It is zero if and only if the graph is disconnected. The number of zero eigenvalues in the eigenvalue set of $L$ equals to the number of connected components in a graph. According to [28], the Fiedler value represented by $\lambda_1$, of a graph, $\Gamma$, can be obtained by the following eigenvalue optimization problem.

$$\begin{aligned}\lambda_1 &= \min\ \mathbf{y}^T L(V, E) \mathbf{y} \\ st.\ \mathbf{y}^T \mathbf{y} &= 1\ and\ \mathbf{y}^T \mathbf{e} = 0\end{aligned} \quad (4)$$

where $\mathbf{y}$ is a vector which does not equal to $\mathbf{e}$, with $\mathbf{e}^T = (1, 1, \ldots, 1)$, and $M^T$ is a transpose to matrix $M$.

The Laplacian matrix of a given graph is defined as follows: Given a graph $\Gamma(V, E)$ without self cycles and multiple links between two nodes, the Laplacian matrix $L$ is calculated by

$$L(V, E) = D(V, E) - A(V, E) \quad (5)$$

where $D(V, E)$ is a diagonal matrix whose diagonal entry contains the degrees for each node. $A(V, E)$ is the adjacency matrix with each entry being a value of zero or one when nodes are connected to each other. In addition, its diagonal is zero because $\Gamma(V, E)$ has no self cycles.

In a network topology graph, $V$ is the set of users, and $E$ is the set of links which support duplex communications, we assume a link exists if and only if bilateral communication between two users is possible.

Because the amplitude of Fiedler value is proportional to the connectivity of networks which is known as the algebraic connectivity, we adopt the algebraic connectivity as our connectivity metric. The smaller the Fiedler value is, the larger the negative impact is onto the networks. Assume $\lambda_i$ is Fiedler value of a graph $\Gamma\backslash\{i\}$ by removing all the incident edges attached to node $i$ in the graph $\Gamma$. Here, we consider a jammer can only turn off one node. These Fiedler values, $\{\lambda_i(\Gamma\backslash\{i\}) | i = 1, 2, \ldots, n\}$, on remaining graphs obtained by removing a different node from the same graph, $\Gamma$, are comparable in terms of graph connectivity although they have different connections on the same number of vertexes. $\lambda_i(\Gamma\backslash\{i\})$ does not relate to the position of node in the graph and nodes' labels. Figure 1 shows an example that the connectivity in different graphs obtained by removing different nodes from the same graph is comparable as long as the number of nodes in remaining graph is the same.

If the jammer aims to reduce network connectivity, Fiedler Value $\{\lambda_i\}$ can be considered as the cost of such adversary's attack. Namely, the jammer's cost of attack to disrupt connectivity is

$$\bar{\lambda}_i = \lambda_i(\Gamma\backslash\{i\}). \quad (6)$$

### 3.2. Cost of jamming throughput attack

If the adversary targets to harm network's throughput, then the total number of throughput for the unaffected network can be considered as a cost of such attack. If the adversary has a selected node *i* for low-power jamming attack, and because this attack also blocks ACK, the total throughput for the rest of the network, or the cost of the throughput jamming attack is given as follows:

$$\bar{\lambda}_i = \sum_{l,j=1, j\neq i, l\neq i}^{n} \ln\left(1 + \frac{h_{l,j}P_{l,j}}{\sigma^2 + \sum_{k=1, k\neq j}^{n} h_{k,j}P_{k,j}}\right). \quad (7)$$

## 4. A STOCHASTIC GAME OF INTRUSION PREVENTION

We consider, as a motivating application, the problem of mitigating an attack directed against an ad hoc network, as depicted in Figure 2. In this scenario, a jammer aims to hurt network by choosing a node to direct interference against, while the network itself aims to reduce the harm this attack has on the network by scanning to detect the harm and ultimately force the adversary into more costly option for conducting its attack.

The category of jammer's attack is fixed in the entire intrusion, which might either be a category of jamming throughput or disrupting connectivity attack. The jammer senses a node which could mostly jeopardize network connectivity through blocking its communication.

In some probability, if the victim node determined by a jammer is also simultaneously scanned by the scanner, because the scanner is present (a jammer can observe presence of authority by only watching or executing some detection techniques which is why he has no intention to perform jamming attack because of the fear of being caught), the jammer switches to the hiding (silent) mode, and if he is not caught, he can continue his attack. However, if the node he chooses is not scanned, the jammer performs an attack. Let $C_h$ be the cost of hiding mode for the jammer corresponding to an applied category of the attack. Let $\alpha$ be the probability to be detected in the hiding mode, and $1-\alpha$ be the probability not to be detected. Thus, the instantaneous cost to the jammer combines the expected hiding cost and cost of network penetration in future if the jammer is not caught. Please note our method can also be applied to hierarchical networks (say, Wifi networks) by assigning more weights to critical nodes, such as access points and cluster heads.

Therefore, we propose the strategies to prevent such attacks in addition to the design of a defense network. Assuming the instantaneous payoff to the legitimate authority equals the instantaneous cost to the jammer. This recursively played zero-sum game $G_\gamma$ can be considered as a single state stochastic game ([29]), which we are going to solve by stationary strategies (i.e., the strategies which do not depend on history and time slot), and it is given as follows:

$$G_\gamma = \begin{matrix} & 1 & 2 & \ldots & n \\ 1 \\ 2 \\ \ldots \\ n \end{matrix} \begin{pmatrix} C_h + \gamma G_\gamma & \bar{\lambda}_2 & \ldots & \bar{\lambda}_n \\ \bar{\lambda}_1 & C_h + \gamma G_\gamma & \ldots & \bar{\lambda}_n \\ \ldots & \ldots & \ldots & \ldots \\ \bar{\lambda}_1 & \bar{\lambda}_2 & \ldots & C_h + \gamma G_\gamma \end{pmatrix}, \quad (8)$$

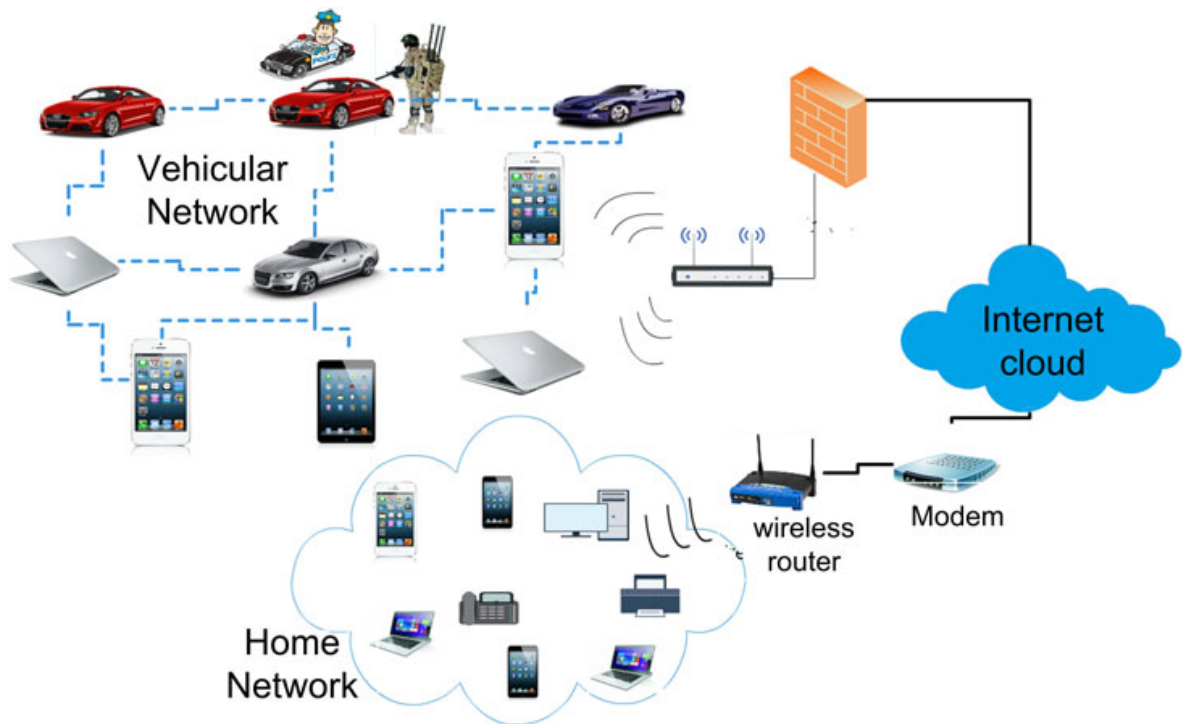

**Figure 2.** A connectivity attack in an ad hoc network.

where rows correspond to the authority's strategies, that is, chosen nodes to scan, and columns correspond to the jammer's strategies, that is, chosen nodes to attack.

Let us describe in details the components of this matrix. Assume the authority has chosen strategy $i$, and the jammer has chosen strategy $j$. If $i \neq j$, node $j$ is jammed successfully, the jammer suffers the instantaneous cost $\bar{\lambda}_j$, and the game is over. If $i = j$ then the jammer switches to the hiding mode paying instantaneous cost $C_h$. With probability $\alpha$, the jammer will be detected, and the game is over. However, if the jammer is not detected, with probability $1 - \delta$, he stops the attempts and exits the game. The game is over. Whereas, with probability $\delta$, the jammer keeps playing the game recursively. Therefore, the instantaneous reward for authority is $\alpha C_h + (1 - \alpha) \left[ C_h + \delta \cdot \text{val}(G_\gamma) \right]$. Then, the conditional probability to keep on the jamming attacks is $\gamma = (1-\alpha)\delta$, and with this probability the game $G$ is played recursively with the expected instantaneous jammer's costs accumulated as $C_h + \gamma G_\gamma$. Because $\gamma < 1$, it can be considered as a discount factor and is the condition that guarantees the convergence of the solution. Here, employing stochastic game tools is quite natural, because the authority and the jammer have opposing objectives, and it is uncertain how persistent the jammer can manage to perform its malicious attack before it is detected. The applications of stochastic games in modeling network security can be found in [30–33] and [34]. Finally, note that the game (8) can be used to model different types of attacks by assigning appropriate content of its parameters. Accordingly, the variable, $\bar{\lambda}_i$, can correspond to either the network's connectivity in a connectivity disruption attack, or the network's throughput in a throughput disruption attack.

Game $G_\gamma$ has a solution in (mixed) stationary strategies, that is, the strategies that are independent of history and current time slot. A (mixed) stationary strategy to the authority is a probability vector $\boldsymbol{p}^T = (p_1, p_2, \ldots, p_n)$, where $p_i$ is the probability to scan node $i$ and $\boldsymbol{e}^T \boldsymbol{p} = 1$. A (mixed) stationary strategy to the jammer is a probability vector $\boldsymbol{q}^T = (q_1, q_2, \ldots, q_n)$, where $q_i$ is the probability to jam node $i$, and $\boldsymbol{e}^T \boldsymbol{q} = 1$. Solution of the game $G_\gamma$ is given as a solution to the Shapley (-Bellmann) equation game [29]:

$$\text{val}(G_\gamma) = \max_{\boldsymbol{p} \geq 0, \boldsymbol{e}^T\boldsymbol{p}=1} \min_{\boldsymbol{q} \geq 0, \boldsymbol{e}^T\boldsymbol{q}=1} \sum_{i=1}^{n} \sum_{j=1}^{n} A_{ij}(\text{val}(G_\gamma)) p_i q_j,$$

$$= \min_{\boldsymbol{q} \geq 0, \boldsymbol{e}^T\boldsymbol{q}=1} \max_{\boldsymbol{p} \geq 0, \boldsymbol{e}^T\boldsymbol{p}=1} \sum_{i=1}^{n} \sum_{j=1}^{n} A_{ij}(\text{val}(G_\gamma)) p_i q_j,$$

(9)

$$A_{ij}(x) = \begin{cases} C_h + \gamma x, & i = j, \\ \bar{\lambda}_j, & i \neq j, \end{cases} \quad (10)$$

and $V_\gamma := \text{val}(G_\gamma)$ is the value of the game, that is, the optimal accumulated cost to the jammer.

Without loss of generality, we can assume that all the nodes have different jamming costs, that is, $\bar{\lambda}_i \neq \bar{\lambda}_j$ for $i \neq j$. Also, let all the nodes are indexed in ascending order by $\bar{\lambda}_i$, that is,

$$\bar{\lambda}_1 < \bar{\lambda}_2 < \ldots < \bar{\lambda}_n. \quad (11)$$

Despite the fact that the stochastic game considered has $n \times n$ instantaneous payoff matrix, we can obtain the solution explicitly from the following theorem given below:

**Theorem 1.** *The stochastic game $G_\gamma$ has an equilibrium in stationary strategies $(\boldsymbol{p}, \boldsymbol{q})$ and the value $V_\gamma$ given as follows:*

*(a)* Let

$$C_h/(1-\gamma) < \bar{\lambda}_1. \quad (12)$$

*Then*

$$V_\gamma = \bar{\lambda}_1,$$

$$p_i \begin{cases} = 0, & i = 1, \\ \geq \frac{\bar{\lambda}_i - \bar{\lambda}_1}{\bar{\lambda}_i - C_h - \gamma \bar{\lambda}_1}, & i \geq 2, \end{cases} \quad (13)$$

$$q_i = \begin{cases} 1, & i = 1, \\ 0, & i \geq 2. \end{cases}$$

*(b)* Let

$$\bar{\lambda}_1 \leq C_h/(1-\gamma) < \lambda_2. \quad (14)$$

*Then*

$$V_\gamma = C_h/(1-\gamma),$$

$$p_i(x) = \begin{cases} 1, & i = 1, \\ 0, & i \geq 2, \end{cases} \quad (15)$$

$$q_i(x) = \begin{cases} 1, & i = 1, \\ 0, & i \geq 2. \end{cases}$$

*(c)* Let

$$\bar{\lambda}_k < C_h/(1-\gamma) \leq \bar{\lambda}_{k+1} \quad (16)$$

*with $\lambda_{n+1} = \infty$, and $m \in [1, k]$ be such that*

$$\varphi_m^{k+1} \leq 1 < \varphi_{m+1}^{k+1}, \quad (17)$$

*with*

$$\varphi_s^{k+1} = \sum_{i=1}^{s} \frac{\bar{\lambda}_s - \bar{\lambda}_i}{C_h + \gamma \bar{\lambda}_s - \bar{\lambda}_i} \text{ for } s \leq k \quad (18)$$

*and $\varphi_{k+1}^{k+1} = \infty$. Note that, by (16), $\varphi_s^{k+1}$ is increasing from zero for $s = 1$ to infinity for $s = k + 1$. Thus, $m$ is uniquely defined by (17).*

*Then,*

$$p_i = \begin{cases} \frac{V_\gamma - \bar{\lambda}_i}{C_h + \gamma V_\gamma - \bar{\lambda}_i}, & i \leq m, \\ 0, & i > m, \end{cases}$$

$$q_i = \begin{cases} \frac{1/(C_h + \gamma V_\gamma - \bar{\lambda}_i)}{\sum_{j=1}^{m} 1/(C_h + \gamma V_\gamma - \bar{\lambda}_j)}, & i \leq m, \\ 0, & i > m, \end{cases} \quad (19)$$

*and $V_\gamma$ is an unique root of the equation*

$$F_m(V_\gamma) := \sum_{i=1}^{m} \frac{V_\gamma - \bar{\lambda}_i}{C_h + \gamma V_\gamma - \bar{\lambda}_i} = 1. \quad (20)$$

*Proof.* First note that $V_\gamma, p$ and $q$ is a solution of Shapley Equation (9) if and only if

$$V_\gamma = v, \quad (21)$$

$$\max \ v,$$
$$\sum_{i=1}^{n} A_{ij}(V_\gamma) p_i \geq v, \ i \in \{1, \ldots, n\}, \quad (22)$$

$p$ is probability vector,

$$\min \ v,$$
$$\sum_{j=1}^{n} A_{ij}(V_\gamma) q_i \leq v, \ j \in \{1, \ldots, n\}, \quad (23)$$

$q$ is probability vector.

Taking into account (10) and the fact that $p$ and $q$ are probability vectors yield that these LP problems (22) and (23) are equivalent to

$$\max \ v,$$
$$(C_h + \gamma V_\gamma - \bar{\lambda}_i) p_i + \bar{\lambda}_i \geq v, \ i \in \{1, \ldots, n\}, \quad (24)$$

$p$ is probability vector,

$$\min \ v,$$
$$(C_h + \gamma V_\gamma - \bar{\lambda}_i) q_i + \sum_{j=1}^{n} \bar{\lambda}_j q_j \leq v, \ j \in \{1, \ldots, n\}, \quad (25)$$

$q$ is probability vector

Then, (21), (24), and (25) imply that $V_\gamma, p$ and $q$ is a solution of Shapley Equation (9) if and only if the following conditions hold:

$$(C + \gamma V_\gamma - \bar{\lambda}_i) q_i + \sum_{j=1}^{n} \bar{\lambda}_j q_j \begin{cases} = V_\gamma, & p_i > 0, \\ \leq V_\gamma, & p_i = 0, \end{cases} \quad (26)$$

$$(C + \gamma V_\gamma - \bar{\lambda}_i) p_i + \bar{\lambda}_i \begin{cases} = V_\gamma, & q_i > 0, \\ \geq V_\gamma, & q_i = 0. \end{cases} \quad (27)$$

Let (12) hold. Then, by (11), (26), and (27), there is no $i$ such that $p_i > 0$ and $q_i > 0$. Also, $q_1 = 1$ and $p_1 = 0$. Substituting them into (26) and (27) implies (a). □

Let (12) do not hold. Then, by (11), (26). and (27), there is a $m$ such that $p_i > 0$ and $q_i > 0$.

$$p_i \begin{cases} > 0, & i \leq m, \\ = 0, & i > m \end{cases} \quad \text{and} \quad q_i \begin{cases} > 0, & i \leq m, \\ = 0, & i > m. \end{cases} \quad (28)$$

Let $m = 1$. Then, by (11), (26), and (27), the condition (14) has to hold, and (b) follows.

Let (16) hold. Note that

$$\max \left\{ \frac{\bar{\lambda}_i - C_h}{\gamma}, \bar{\lambda}_i \right\} = \begin{cases} \frac{\bar{\lambda}_i - C_h}{\gamma}, & \bar{\lambda}_i \geq \frac{C_h}{1-\gamma}, \\ \bar{\lambda}_i, & \bar{\lambda}_i \leq \frac{C_h}{1-\gamma}. \end{cases} \quad (29)$$

Because $m > 1$, by (11), (26), (27), and (28) $p$ and $q$ have to have the form given by (19). Then, because sums of the components of vector $p$ equals 1, $V$ has to be given as a root of the Equation (20). It is only left to show that this equation has a unique root. Let $m < k$. Then, by (11) and (29), $F_m$ is increasing in $[\bar{\lambda}_m, \bar{\lambda}_{m+1}]$ such that, by (17), $F_m(\bar{\lambda}_m) = \varphi_m^{k+1} \leq 1 < \varphi_{m+1}^{k+1} = F_m(\bar{\lambda}_{m+1})$. Thus, $V$ is uniquely defined. Let $m = k$. Then, by (11) and (29), $F_k$ is increasing in $[\bar{\lambda}_k, (\bar{\lambda}_k - C)/\gamma]$, and $F_k((\bar{\lambda}_k - C)/\gamma) > 1$, and (c) follows.

Theorem 1 allows to observe some interesting properties of the solution.

If hiding cost $C_h$ is too big, namely, $C_h \geq \bar{\lambda}_n$, then all the nodes will be under attack, and thus, have to be scanned, that is, $p_i > 0$ and $q_i > 0$ for any $i$, and the value of the game is the unique root of the equation $F_n(V_\gamma) = 1$. Also, the value $V_\gamma$ of the game is increasing on $C_h$ and $\gamma$ where including $\gamma = 1$.

Because the game $G_0$ is one time slot game, it is just a matrix game. Its solution is given in the following theorem in the closed form.

**Theorem 2.** *One time slot matrix game $G_0$, which is the limit of the stochastic game $G_\gamma$ for $\gamma$ tends to zero, has value $V_0 = V(C_h)$ and the equilibrium strategies $p$ and $q$ given as follows:*

(a) *Let*

$$C_h < \bar{\lambda}_1. \quad (30)$$

*Then*

$$V(C_h) = \bar{\lambda}_1,$$

$$p_i \begin{cases} = 0, & i = 1, \\ \geq \frac{\bar{\lambda}_i - \bar{\lambda}_1}{\bar{\lambda}_i - C_h}, & i \geq 2, \end{cases} \quad (31)$$

$$q_i(x) = \begin{cases} 1, & i = 1, \\ 0, & i \geq 2. \end{cases}$$

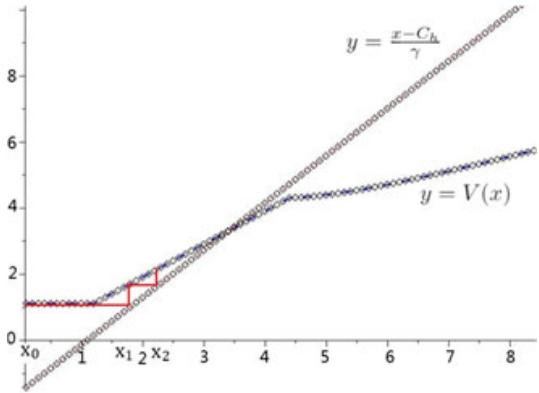

**Figure 3.** Convergence of iterative procedure for $n = 5$, $\bar{\lambda} = (1.11, 4.31, 6.12, 8.31, 9.11)$, $C_h = 1$ and $\gamma = 0.7$.

*(b) Let*
$$\bar{\lambda}_1 \leq C_h < \bar{\lambda}_2. \tag{32}$$

*Then*
$$V(C_h) = C_h,$$
$$p_i = \begin{cases} 1, & i = 1, \\ 0, & i \geq 2, \end{cases} \tag{33}$$
$$q_i = \begin{cases} 1, & i = 1, \\ 0, & i \geq 2. \end{cases}$$

*(c) Let*
$$\bar{\lambda}_k < C_h \leq \bar{\lambda}_{k+1} \tag{34}$$

*and $m$ be given by (17). Then,*

$$V(C_h) = \frac{1 + \sum_{j=1}^{m} \bar{\lambda}_j/(C_h - \bar{\lambda}_j)}{\sum_{j=1}^{m} 1/(C_h - \bar{\lambda}_j)},$$

$$p_i = \begin{cases} \frac{V(C_h) - \bar{\lambda}_i}{C_h - \bar{\lambda}_i}, & i \leq m, \\ 0, & i > m, \end{cases} \tag{35}$$

$$q_i = \begin{cases} \frac{1/(C_h - \bar{\lambda}_i)}{\sum_{j=1}^{m} 1/(C_h - \bar{\lambda}_j)}, & i \leq m, \\ 0, & i > m. \end{cases}$$

Theorem 2 also allows to suggest two procedures to find the value of the stochastic game.

**Theorem 3.** *The value of the stochastic game $G_\gamma$ is given as follows*

$$V_\gamma = \frac{x - C_h}{\gamma},$$

*where $x \geq C_h$ is a unique root of the equation*

$$\frac{x - C_h}{\gamma} = V(x). \tag{36}$$

*The unique root of (36) can be found by*

*(a) iterative procedure $x_0 = C_h$, $x_{i+1} = \gamma V(x_i) + C_h$, $i = 0, 1, \ldots$ until $|x_{i+1} - x_i| \leq \epsilon$ with $\epsilon$ is tolerance.*
*(b) bisection method because $(x - C_h)/\gamma - V(x) = -C_h/\gamma - v(0) < 0$ for $x = 0$ and $(x - C_h)/\gamma - v(x) > 0$ for enough large x.*

Figure 3 illustrates convergence of iterative procedure to the equilibrium point.

## 5. MAXMIN SELECTION OF SCANNING STRATEGY

In reality, the jammer can deteriorate network performance in many aspects such as reducing either its connectivity or throughput or secrecy communication. Motivated by these different categories of malicious activity, the jammer could vary the corresponding optimal strategies. However, the authority might have no knowledge of the jammer's motivation for an attack, and so about the strategy employed. The authority might only know the set of all possible motivations and the corresponding optimal strategies used by the jammer.

Under this situation, the need for comparing these strategies arises because they aim to achieve different metrics. Say, connectivity is a metric for the strategies which aim to jeopardize network connectivity, whereas, throughput is a metric for the strategies which aim to harm the throughput. However, in spite of differences in metrics, the ultimate goal of a jammer is to speed up the process of completing the attack because long time commitment involves more cost. Thus, the expected time of successful attack can be considered as a common metric for all the categories of malicious activity, where the authority wants to maximize this metric while the jammer aims to minimize it. If we assume the rival chooses a specific category and he follows the category over time before completing the attack, then the expected jamming time, $T$, before a successful attack appears, can be represented as following:

$$T(\boldsymbol{p}, \boldsymbol{q}) = \sum_{t=1}^{\infty} t \left[ \left( \sum_{i=1}^{n} \gamma p_i q_i \right)^{t-1} \left( 1 - \sum_{i=1}^{n} \gamma p_i q_i \right) \right] \tag{37}$$

$$= \frac{1}{1 - \gamma \boldsymbol{p}^T \boldsymbol{q}}. \tag{38}$$

Where $\boldsymbol{q}$ is a probability vector which represents a category of strategies employed by the jammer. $\boldsymbol{p}$ is a probability vector which represents a category of strategy applied by the authority to scan the attack. Thus, these strategies

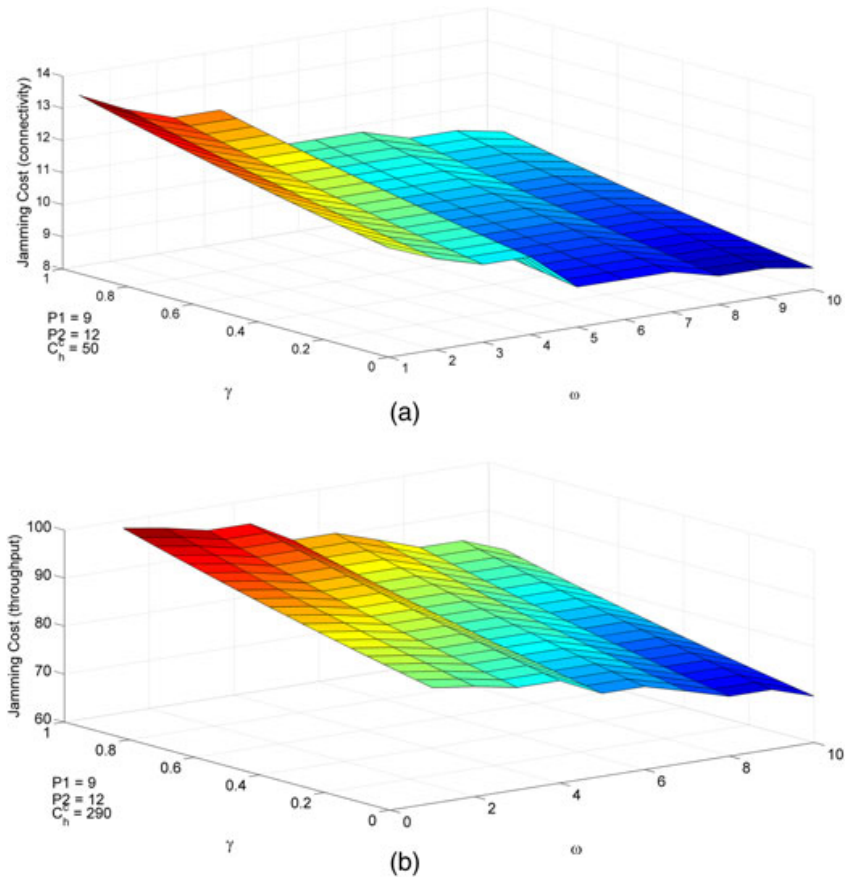

**Figure 4.** Stochastic game: accumulated connectivity (up) and throughput (down) costs as function of receiver's signal to interference and noise ratio threshold, $\omega$, and probability of re-playing the game, $\gamma$.

depend on the category of malicious activity chosen by the jammer.

Although suggested approach might be applied to any category of an attack, to get insight of the problem, we focus only on two which are also the most important metrics for network performance, namely, network's connectivity and network's throughput. We denote these metrics (connectivity, throughput) by the symbols, "$c$" and "$t$". The optimal strategy pair, $(\boldsymbol{p}_c, \boldsymbol{q}_c)$ for dealing with attack aiming to destroy connectivity, was found in the previous section. The optimal strategies $(\boldsymbol{p}_t, \boldsymbol{q}_t)$ for dealing with attack aiming to harm throughput, can be found by substituting connectivity cost $\bar{\lambda}_i$ with total remaining throughput expressed in Equation (7) into matrix (8).

The authority wants to maximize the jammer's attacking time in order to force the jammer to make his attack more expensive. Whereas, the jammer wants to minimize it. The authority does not know what category of the attack the jammer intends to follow. The jammer does not know versus what category of the attack the authority intends to build up his defense. Thus, the rival faces with a dilemma of choosing the proper strategies. This dilemma can be described by the following zero-sum $2 \times 2$ matrix game

$$D = \begin{matrix} c \\ t \end{matrix} \begin{pmatrix} T(\boldsymbol{p}_c, \boldsymbol{q}_c) & T(\boldsymbol{p}_c, \boldsymbol{q}_t) \\ T(\boldsymbol{p}_t, \boldsymbol{q}_c) & T(\boldsymbol{p}_t, \boldsymbol{q}_t) \end{pmatrix},$$

where rows correspond to the authority's strategies, that is, choosing attack's category to response, and columns correspond to the jammer's strategies, that is, choosing attack's category.

This matrix game has an equilibrium ([35]) either in pure strategies, that is, when the rival selects a specific one, or in mixed strategies, when the rival randomizes his selection. Because the game is zero-sum, then the authority's equilibrium strategy is also his maxmin strategy, that is, it is the best response strategy for the most dangerous adversary's attack. This result is given in the following two propositions.

**Proposition 1.** *The game has an equilibrium in (pure) strategies if and only if the following conditions hold:*

(1) *If* $\boldsymbol{p}_t^T \boldsymbol{q}_c \leq \boldsymbol{p}_c^T \boldsymbol{q}_c \leq \boldsymbol{p}_c^T \boldsymbol{q}_t$ *then* $(c, c)$ *is an equilibrium,*
(2) *If* $\boldsymbol{p}_c^T \boldsymbol{q}_c \leq \boldsymbol{p}_t^T q_c \leq \boldsymbol{p}_t^T \boldsymbol{q}_t$ *then* $(t, c)$ *is an equilibrium,*
(3) *If* $\boldsymbol{p}_t^T \boldsymbol{q}_t \leq \boldsymbol{p}_c^T \boldsymbol{q}_t \leq \boldsymbol{p}_c^T \boldsymbol{q}_c$ *then* $(c, t)$ *is an equilibrium,*
(4) *If* $\boldsymbol{p}_c^T \boldsymbol{q}_t \leq \boldsymbol{p}_t^T \boldsymbol{q}_t \leq \boldsymbol{p}_t^T \boldsymbol{q}_c$ *then* $(t, t)$ *is an equilibrium.*

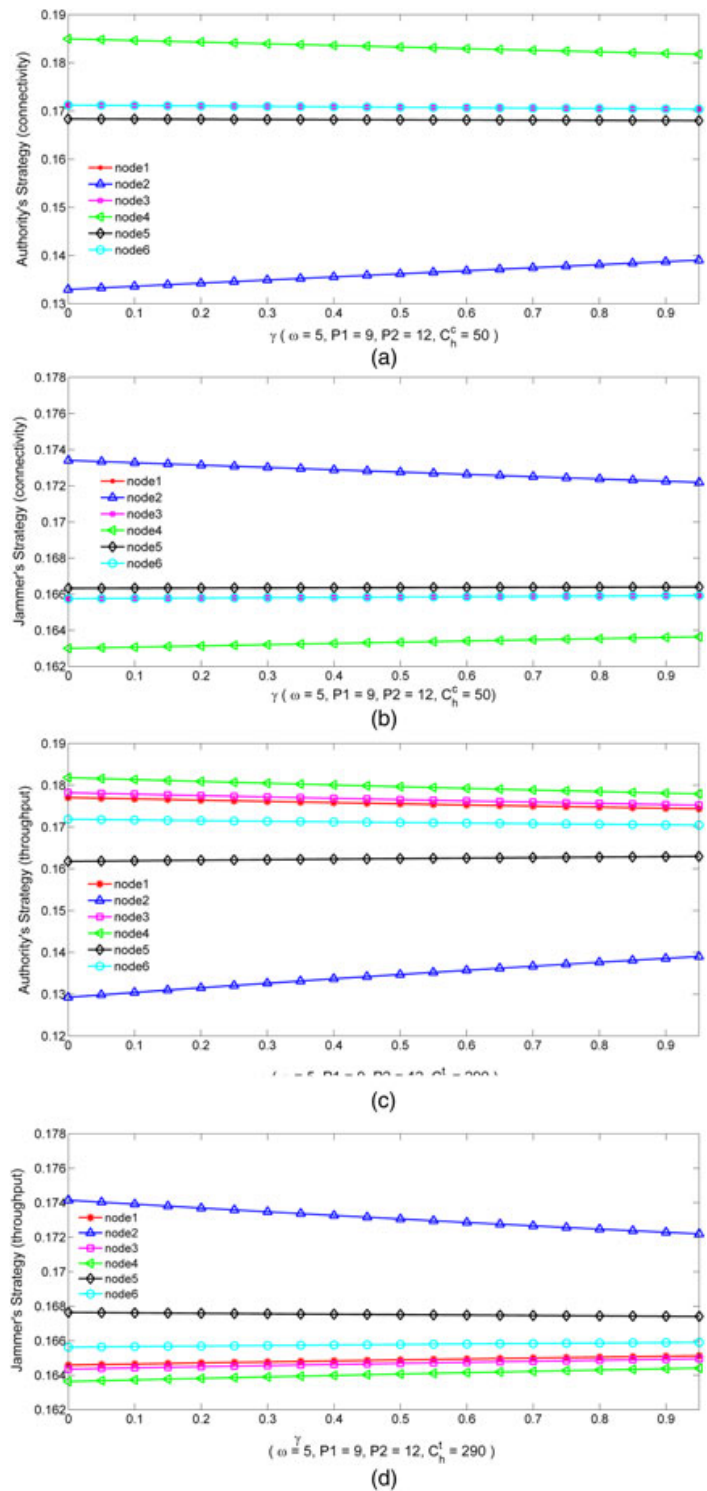

**Figure 5.** Stochastic game: (a) the authority's strategy for connectivity game, (b) the jammer's strategy for connectivity game, (c) the authority strategy for throughput jamming game and (d) the jammer's strategy for throughput jamming game as functions of probability for the game to be continued, $\gamma$.

**Proposition 2.** *If there is no equilibrium in pure strategies, the rival applies the randomized strategies. Namely, with probability, $x_c$ ($x_t$), the authority should defend against "c" ("t") attack's category, and with probability, $y_c$ ($y_t$), the jammer applies strategy corresponding to "c" ("t") attack's category, where*

$$x_c = \frac{T(\mathbf{p}_t,\mathbf{q}_t) - T(\mathbf{p}_t,\mathbf{q}_c)}{T(\mathbf{p}_c,\mathbf{q}_c) + T(\mathbf{p}_t,\mathbf{q}_t) - T(\mathbf{p}_c,\mathbf{q}_t) - T(\mathbf{p}_t,\mathbf{q}_c)},$$
$$x_t = 1 - x_c,$$
$$y_c = \frac{T(\mathbf{p}_t,\mathbf{q}_t) - T(\mathbf{p}_c,\mathbf{q}_t)}{T(\mathbf{p}_c,\mathbf{q}_c) + T(\mathbf{p}_t,\mathbf{q}_t) - T(\mathbf{p}_c,\mathbf{q}_t) - T(\mathbf{p}_t,\mathbf{q}_c)}, \quad (39)$$
$$y_t = 1 - y_c.$$

## 6. SIMULATION

In this section, numerical results are given to illustrate the impact of network parameters, such as transmission power of nodes and SINR's threshold, on maintaining the communication links. In simulation setting, the network consists of six nodes, that is, $n = 6$ with background noise $\sigma^2$ equals to one. The authority scans the network to prevent malicious activity, and he does not participate in packet transmissions. The channel gain matrix, $\mathbf{h}$, are randomly generated and given as follows:

$$\mathbf{h} = \begin{bmatrix} 0 & 0.3128 & 1.1790 & 1.6488 & 1.6335 & 0.8458 \\ 0.3128 & 0 & 0.4524 & 1.9653 & 0.5215 & 0.1885 \\ 1.1790 & 0.4524 & 0 & 1.4605 & 1.1887 & 1.1970 \\ 1.6488 & 1.9653 & 1.4605 & 0 & 0.0450 & 0.9418 \\ 1.6355 & 0.5215 & 1.1887 & 0.0450 & 0 & 1.3919 \\ 0.8458 & 0.1885 & 1.1970 & 0.9418 & 1.3919 & 0 \end{bmatrix}.$$

For each node as a transmission protocol, we consider the uniform power allocation that transmits the same power signals to its neighbors, but each node, of course, can have different total power levels. Here, we consider that transmission powers ($P1$ and $P2$) of node 1 and 2 vary from 0 to 20. Transmission power of node 3, $P3$, is 11. $P4$ is 10. $P5$ is 9. $P6$ is 8. Note that, the protocol of uniformly allocating transmission power is proved to be optimal for independent and identically distributed Gaussian channels[36,37]. Thus, given $\mathbf{h}$, $P$, $\sigma^2$ and receiving threshold, they already define a network's topology.

Figure 4 illustrates the accumulated connectivity and throughput costs, that is, the value $V_\gamma$ of the game $G_\gamma$, as function of receiver's SINR threshold, $\omega$, and probability of re-playing the game, $\gamma$. It shows that the accumulated jamming cost decreases with increasing threshold because the larger value of threshold assumes smaller effort to break communication links, and so smaller efforts to harm networks' connectivity. Also, increasing probability of re-playing the game, $\gamma$, yields in raising the value of the game, and so the jamming cost because longer duration of the game calls for the growth on the accumulated hiding cost. The smallest jamming cost is achieved when $\gamma = 0$, which is when the jammer can manage to perform only one time shot attack.

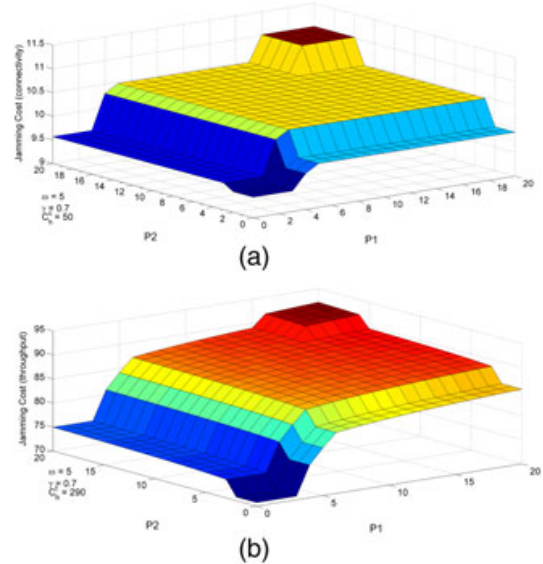

**Figure 6.** Stochastic game: accumulated connectivity (up) and throughput (down) costs as functions of transmission powers of node 1 and node 2.

Figure 5 illustrates the authority's and the jammer's strategies for connectivity and throughput jamming game as function of probability for the game are continuous. It shows how the jammer tries to avoid being detected by the authority, in order to perform a successful attack.

Figure 6 illustrates the relations between the jamming cost and transmission power. These relations are piece-wise continuous with jumps happen in between. Increasing transmission power causes adding new communication links into the network topology. This yields into increasing Fiedler value and throughput by jumps. Thus, it produces the leap in the value of the game. Because the interference is not considered in the simulation for showing an obvious tendency without fluctuation, the value of connectivity game is piece-wise constant on transmission power while the value of throughput jamming game is piece-wise continuous.

Figure 7(a) illustrates the probability that the authority intends to deal with an attack aiming to disrupt connectivity as function of transmission power of node 2. It shows that bigger probability $\gamma$ assumes smaller transmission power in order to switch to mixed strategy. Also, the authority's strategy on maintaining connectivity is non-increasing on the transmission power. Figure 7(b) illustrates the expected duration of the game as function on transmission power of node 2. On the same reason as one for the value of the game, namely, changing in the network's topology due to adding new links, the expected duration of the game are piece-wise continuous function on the transmission power.

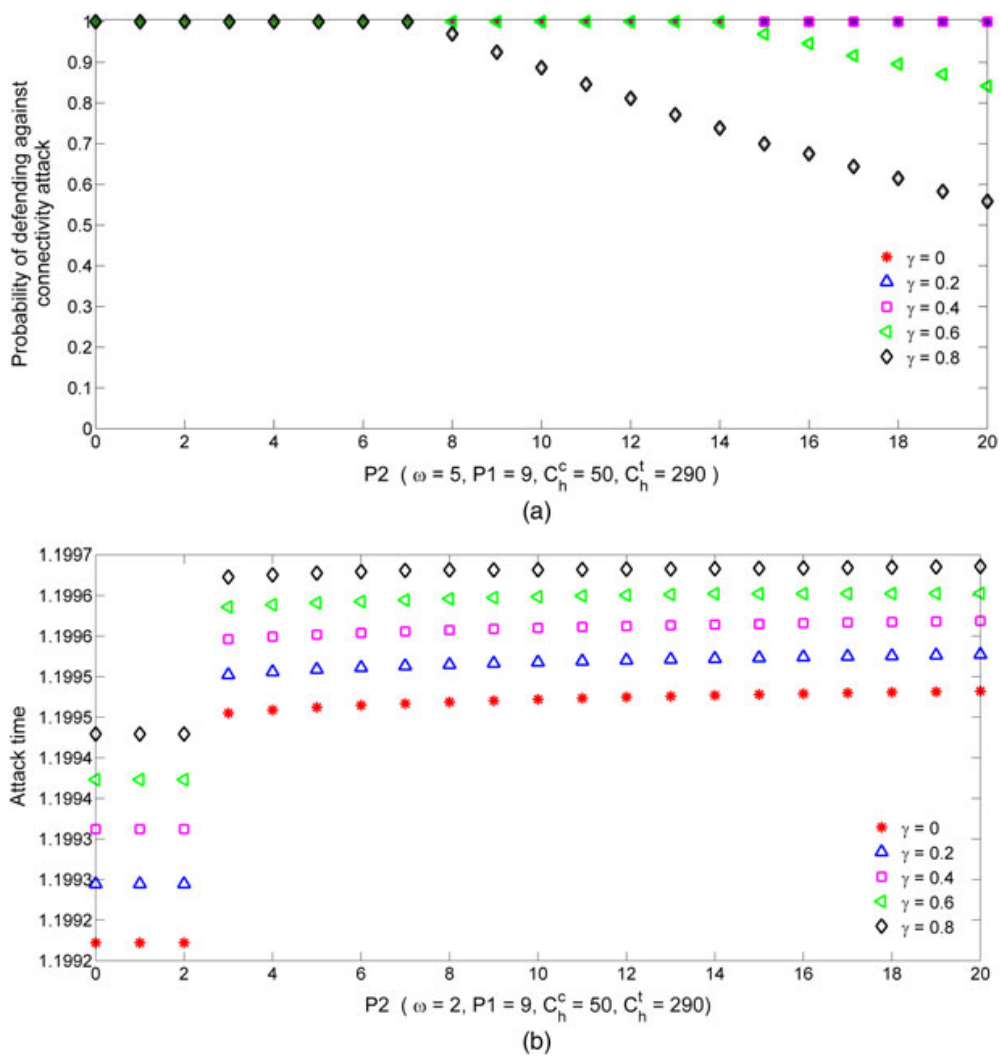

**Figure 7.** Maxmin selection of scanning strategy: (a) the probability that the authority intends to deal with attack aiming to disrupt connectivity, and (b) the expected durations of the game as function on transmission power of node 2.

## 7. CONCLUSIONS

In this paper, costs for jamming attack on connectivity and throughput are introduced. Then, a uniform stochastic game with network scanning to prevent jamming attack is suggested and solved explicitly. Because of different aims, the jammer might choose different categories of attacks, for example, connectivity and throughput. Comparing such incomparable attacks is a challenge. To deal with this issue, an approach is suggested to compare them by duration of attack instead of the damage they bring to. Game theoretical model for such comparison is suggested, and optimal strategies are proposed. Results for numerical evaluation of the optimal solutions and their dependence on network's characteristics are supplied. A goal of our future work is to investigate more sophisticate behavior of the adversary where he can switch between different categories of malicious activities based on the archived result of attack, and to incorporate mechanism of learning in the authority strategy based on the accumulated results of scanning.

## REFERENCES

1. Lunt TF. A survey of intrusion detection techniques. *Computers & Security* 1993; **12**(4): 405–418.
2. Mukherjee B, Heberlein LT, Levitt KN. Network intrusion detection. *IEEE Network* 1994; **8**(3): 26–41.
3. Garnaev A, Trappe W, Kung C-T. Dependence of optimal monitoring strategy on the application to be protected. *Proc. IEEE Global Communications Conference (GLOBECOM '12)*, Anaheim, CA, USA, 2012; 1054–1059.
4. Garnaev A, Trappe W. Bandwidth scanning involving a Bayesian approach to adapting the belief of

an adversary's presence. *Proc. IEEE Conference on Communications and Network Security (CNS)*, San Francisco, CA, USA, 2014; 35–43.

5. Garnaev A, Liu Y, Trappe W. Anti-jamming strategy versus a low-power jamming attack when intelligence of adversary's attack type is unknown. *IEEE Transactions on Signal and Information Processing over Networks* 2016; **2**(1): 49–56.

6. Lee W, Stolfo SJ. Data mining approaches for intrusion detection. *Proc. the 7th Conference on USENIX Security Symposium*, vol. 7, USENIX Association, Berkeley, CA, USA, 1998; 6–6.

7. Lee W, Stolfo SJ, Mok KW. A data mining framework for building intrusion detection models. *Proc. the 1999 IEEE Symposium on Security and Privacy*, Oakland, CA, USA, 1999; 120–132.

8. Pelechrinis K, Koutsopoulos I, Broustis I, Krishnamurthy SV. Lightweight jammer localization in wireless networks: system design and implementation. *Proc. IEEE Global Telecommunications Conference (GLOBECOM '09)*, Honolulu, HI, USA, 2009; 1–6.

9. Liu Y, Trappe W. Jammer forensics: localization in peer to peer networks based on q-learning. *Proceedings of IEEE International Conference on Acoustics, Speech and Signal Processing (ICASSP)*, South Brisbane, QLD, Australia, 2015; 1737–1741.

10. Liu Y, Garnaev A, Trappe W. Maintaining throughput network connectivity in ad hoc networks. *Proc. IEEE International Conference on Acoustics, Speech and Signal Processing (ICASSP)*, Shanghai, China, 2016.

11. Chung FR. Spectral graph theory. *American Mathematical Society* 1997; **92**.

12. Albert R, Jeong H, Barabási AL. Error and attack tolerance of complex networks. *Nature* 2000; **406**(6794): 378–382.

13. Newman ME. The structure and function of complex networks. *SIAM Review* 2003; **45**(2): 167–256.

14. Albert R, Barabási AL. Statistical mechanics of complex networks. *Reviews of modern physics* 2002; **74**(1): 47.

15. Haythornthwaite C. Social networks and internet connectivity effects. *Information, Community & Society* 2005; **8**(2): 125–147.

16. Hummon NP, Dereian P. Connectivity in a citation network: the development of DNA theory. *Social Networks* 1989; **11**(1): 39–63.

17. Walters JP, Liang Z, Shi W, Chaudhary V. Wireless sensor network security: a survey. *Security in Distributed, Grid, Mobile, and Pervasive Computing* 2007; **1**: 367.

18. Karlof C, Wagner D. Secure routing in wireless sensor networks: attacks and countermeasures. *Ad hoc networks* 2003; **1**(2): 293–315.

19. Yang H, Luo H, Ye F, Lu S, Zhang L. Security in mobile ad hoc networks: challenges and solutions. *IEEE Wireless Communications* 2004; **11**(1): 38–47.

20. Roy S, Ellis C, Shiva S, Dasgupta D, Shandilya V, Wu Q. A survey of game theory as applied to network security. *Proc. IEEE the 43rd Hawaii International Conference on System Sciences (HICSS)*, Honolulu, HI, USA, 2010; 1–10.

21. Manshaei MH, Zhu Q, Alpcan T, Bacșar T, Hubaux JP. Game theory meets network security and privacy. *ACM Computing Surveys (CSUR)* 2013; **45**(3): 25.

22. Gueye A, Walrand JC, Anantharam V. Design of network topology in an adversarial environment. *Proc. the First International Conference on Decision and Game Theory for Security (GameSec 2010)*, Springer, Berlin, Germany, 2010; 1–20.

23. Lindelauf R, Blankers I. Key player identification: a note on weighted connectivity games and the shapley value. *Proc. IEEE International Conference on Advances in Social Networks Analysis and Mining (ASONAM)*, Odense, Denmark, 2010; 356–359.

24. Theodorakopoulos G, Baras JS. A game for ad hoc network connectivity in the presence of malicious users. *Proc. IEEE Global Telecommunications Conference(GLOBECOM '06)*, San Francisco, CA, USA, 2006; 1–5.

25. Zhang Z, Wu J, Deng J, Qiu M. Jamming ack attack to wireless networks and a mitigation approach. *Proc. IEEE Global Telecommunications Conference (GLOBECOM '08)*, New Orleans, LO, USA, 2008; 1–5.

26. Attar A, Tang H, Vasilakos AV, Yu FR, Leung VCM. A survey of security challenges in cognitive radio networks: solutions and future research directions. *Proceedings of the IEEE* 2012; **100**: 3172–3186.

27. Mohar B. Laplace eigenvalues of graphsa survey. *Discrete mathematics* 1992; **109**(1): 171–183.

28. Fiedler M. Algebraic connectivity of graphs. *Czechoslovak Mathematical Journal* 1973; **23**(2): 298–305.

29. Neyman A, Sorin S. *Stochastic Games and Applications*, Vol. 570. Nato Science Series C: Springer Netherlands, 2003.

30. Nguyen KC, Alpcan T, Bașar T. Stochastic games for security in networks with interdependent nodes. *Proc. IEEE International Conference on Game Theory for Networks (GameNets' 09)*, Istanbul, Turkey, 2009; 697–703.

31. Wang B, Wu Y, Liu K, Clancy TC. An anti-jamming stochastic game for cognitive radio networks. *IEEE Journal on Selected Areas in Communications* 2011; **29**(4): 877–889.

32. Garnaev A, Trappe W. Stationary equilibrium strategies for bandwidth scanning. *Proc. the 6th International Workshop on Multiple Access Communcations - Volume 8310*, Springer, New York, NY, USA, 2013; 168–183.
33. Garnaev A, Trappe W. Anti-jamming strategies: a stochastic game approach. *Proc. the 6th International Conference on Mobile Networks and Management*, Springer, Würzburg, Germany, 2015; 230–243.
34. Calinescu G, Kapoor S, Qiao K, Shin J. Stochastic strategic routing reduces attack effects. *Proc. IEEE Global Telecommunications Conference (GLOBECOM '11)*, Houston, TX, USA, 2011; 1–5.
35. Nisan N, Roughgarden T, Tardos E, Vazirani VV. *Algorithmic Game Theory*, Vol. 1. Cambridge University Press Cambridge, 2007.
36. Telatar IE. et al. Capacity of multi-antenna Gaussian channels. *European Transactions on Telecommunications* 1999; **10**(6): 585–595.
37. Rhee W, Cioffi JM. Ergodic capacity of multi-antenna Gaussian multiple-access channels. *Proc. IEEE Conference Record of the Thirty-Fifth Asilomar Conference on Signals*, Pacific Grove, CA, USA, 2001; 507–512.